\documentstyle[11pt,newpasp,twoside,epsf]{article} 
\def\edcomment#1{\iffalse\marginpar{\raggedright\sl#1\/}\else\relax\fi} 
\reversemarginpar 

\begin{document} 
\title{Cluster formation in the LMC}

\author{A. Vallenari, 
A. Moretti, E.Held, L.Rizzi, D.Bettoni} 
\affil{Padova Observatory, Vicolo Osservatorio 5, Padova} 

\begin{abstract} 
In this work we present preliminary results concerning the properties of
about 50 clusters and associations located in a densely populated
region at the Southern edge of the supershell LMC~4. 
The ages of the clusters and associations
are derived. The cluster formation rate peaks at about 10-20 Myr ago corresponding to the age of the formation of LMC~4. Objects younger than 10 Myr are located close to CO clouds, indicating a possible trigger of the recent star formation by the interaction of the supershell with the interstellar medium.
 Isophotal analysis has been done for 50\% of the clusters. 
A large number of objects present isophotal distortions commonly interpreted
as sign of interaction. Cluster multiplets are coeval within 20 Myr, suggesting a common origin. This indicates that clusters possibly form in large groups.
Our findings are in agreement with the unified view of multi-scale star formation, where a size-duration correlation is expected.
\end{abstract}


The LMC/SMC  are of fundamental importance to understand
the  star formation process.  In particular  the presence of
super-giant shells due to stellar winds and supernovae 
allow us to study the secondary and self propagating star formation
in galaxies.
In this work we present preliminary results concerning a large
region centered at the Southern rim of the supershell LMC~4 ($\alpha=05:32:19,
\delta=-67:31:18$ at J2000) where about 50 clusters and associations
are located. CO clouds superposed on the H$\alpha$ emission  have been detected
 at the
South-Eastern side of the observed field (Yamaguchi et al 2001; Mizuno et al 2000). 
The star formation history of both clusters and field stars can provide information about the formation of the super-giant shell LMC~4 and its interactions with the interstellar medium.
B,V,I images have been taken using the ESO 2.2m
telescope equipped with the WFI Imager, covering 30$\prime \times$ 30$\prime$
about. More than 100000 stars are found down to V$\sim 25$.
We derive the cluster/association age distribution from the analysis
of the Colour-Magnitude diagrams, once that the field contamination is
statistically removed.  The cluster formation rate presents two major peaks,
 namely at 10 and 100 Myr.  Minor episodes are found from 60 to 10 Myr ago.
 Finally, about 10\% of the clusters are younger than
10 Myr.
Comparing our age determination with the CO cloud Catalog by Mizuno et al (2001) we find that
while the youngest objects (age less than 10 Myr) are mostly located close to the  H$\alpha$ and CO emission,
the oldest generation of clusters seems to be more evenly distributed.
This is in agreement with the suggestion by Yamaguchi et al (2001) that
the star formation is triggered by the compression of the super-giant shell
on the CO clouds having a lifetime of a few million of years.
 
The star formation history of the field population presents recent episodes of star formation 
aged of 10, 15, 25,100 Myr. A population of a few Gyrs of 
field stars belonging to the
red clump of He-burning objects is found to be uniformly distributed, being part of the
pre-existing generation. Main sequence stars in the age range 20-500 Myr
are showing a  structure ($\sim$ 300 pc) involving a large fraction of the clusters. An age gradient seems to be present in the field population from the inside  towards the external rim of the studied region of the LMC~4.
 The star formation episode aged of 10-20 Myr found either
in the field and in the cluster formation rate is in agreement with the
epoch of the formation of  LMC~4 estimated to happen
 10-20 Myr ago (Vallenari, Bomans \& de Boer 1993; Braun et al 1997).

Isophotal analysis has been done for  about 50\% of the objects classified as
clusters. To remove the field and sky contamination the background outside 3 cluster diameters is fitted and subtracted. Radial profiles are derived first smoothing
the images with a Gaussian filter and then masking the bright stars. Distortions on the isophotal profiles such as tails, bridges, common envelopes are interpreted as sign of interaction between the clusters following
de Oliveira, Dottori \& Bica (1998) and Leon, Bergond \& Vallenari (1999).  To avoid the effect of the noise due to bright stars, we quantify the
asymmetry in the isophote distribution by means of the parameter $\delta_c=
\Delta R/R$ where $\Delta R = ((Xc-Xe)^2+(Yc-Ye)^2)^{1/2}$ following Vallenari et al (1998). R is the radius
of the isophote having the maximum value of the shift and Xe and Ye are the corresponding coordinates of the isophote centers. (Xc,Yc) is the center of the cluster. Unperturbed objects are characterized by   $\delta_c < 0.20$ as already
determined by Vallenari, Bettoni \& Chiosi (1998). Preliminary analysis
shows that in our sample more than 80\% of the analyzed 
objects presents some signs
of distortion having $\delta_c$ in the range 0.20-0.80. Cluster pairs and multiplets are coeval within 20 Myr indicating a common origin. This is in agreement
with the unified view of multi-scale star formation: the turbulence effects
produce hierarchical structures in the gas ensuring that the triggering time scales with the size of the region
(Elmegreen 2002).
Clusters possibly form in large groups with high formation efficiency. Multiplets are expected to merge
into pairs or single objects on a time scale of the order of 100 Myr (de Oliveira et al 2000).




\section{References} 
Braun, J. M., Bomans, D.J., Will, J-M, \& de Boer, K.S., 1997, \aap, 328, 167
Elmegreen B.G., 2002, astro-ph/0207114 \\
de Oliveira M.R, Dottori H., \& Bica E., 1998, \mnras, 295, 921\\
de Oliveira M.R., Dutra C.M., Bica E., \& Dottori H., 2000, \aaps, 146, 57\\
Leon S., Bergond G., \& Vallenari A., 1999, \aap, 344, 450 \\
Mizuno N, Yamaguchi, R., Mizuno, A., Rubio, M., Abe, R., Saito, H., Onishi, T., Yonekura, Y.,Yamaguchi, N., Ogawa, H., Fukui, Y.,
2001,\pasj, 53, 971 \\
Vallenari A., Bettoni D., \& Chiosi C., 1998, \aap, 331, 506\\
Yamaguchi R., Mizuno N., Onishi T., Mizuno A., \& Fukui Y., 2001, \pasj, 53, 959\\
Vallenari A., Bomans, D.J., \& de Boer, K.S., 1993, \aap, 268, 137\\



\end{document}